\documentstyle[prl,aps,twocolumn]{revtex}
\begin{document}
\twocolumn[\hsize\textwidth\columnwidth\hsize\csname@twocolumnfalse%
\endcsname
\title{On phase transitions in two-dimensional disordered systems}
\author{Paul Fendley and Robert M.\ Konik}
\address{Physics Department, University of Virginia,
Charlottesville VA 22901}
\date{May 2000}
\maketitle
\begin{abstract}

The low-energy limits of models with disorder are frequently described
by sigma models. In two dimensions, most sigma models admit either a
Wess-Zumino-Witten term or a theta term.  When such a term is present
the model can have a stable critical point with gapless
excitations. We describe how such a critical point appears, in
particular in two-dimensional superconductors with disorder. The
presence of such terms is required by the underlying (anomalous)
symmetries of the original electron model. This indicates that the
usual symmetry classes of disordered systems in two dimensions can
be further refined. Conversely, our results also indicate that models
previously thought to be in different universality classes are in fact
the same once the appropriate extra terms are included.
\end{abstract}
\pacs{PACS numbers: ???}  ]  
\bigskip

Understanding the effects of disorder is one of the central problems
of condensed matter physics.  It is natural to focus upon the phase
transitions these systems undergo.  In this paper, we will describe
several ways phase transitions and the corresponding gapless modes
arise in disordered models in two dimensions.

The low-energy degrees of freedom for non-interacting fermions with
disorder can frequently be described by a sigma model in the replica
formulation \cite{SW}.  A $G/H$ sigma model arises when the vacuum
manifold of a field is $H$, a subgroup of the global symmetry group
$G$.  The low-energy modes of this model then take values in the coset
$G/H$.  Such sigma models in two dimensions do not generically have
gapless degrees of freedom. For finite number of replicas, this is a
result of the Mermin-Wagner-Coleman theorem. The sigma models are
usually gapped because the space $G/H$ is curved.  Expanding the
action in powers of the curvature gives a mass scale: this is the
well-known story of asymptotic freedom and dimensional
transmutation. At specific values of $N$, the curvature and hence the
beta function can vanish \cite{Gade}, but in this paper we will be
concerned with the general situation.

Despite the fact that sigma models are usually gapped, there are many
non-trivial fixed points in two dimensions.  One can add to the sigma
model action terms which dramatically change the low-energy physics.
One famous example is the topological theta term.  For example, the
model with $G=O(3)$ and $H=O(2)$ has a gap \cite{ZZ}
unless one adds a theta term to the action. This $O(3)$-invariant term
is a total derivative, so it depends only on the fields at spacetime
infinity and has no effect on perturbation theory around the unstable
Gaussian fixed point. Explicitly, the theta term takes the form $i
n\theta$, where $n$ is an integer counting how many times a field
configuration winds around the $O(3)/O(2)$ two-sphere. As has been
discussed in the context of the Heisenberg spin-chain, at $\theta=\pi$
there is a non-trivial low-energy fixed point
\cite{Haldane,ZZ}. It is unstable when $\theta$ is not $\pi$, so
tuning $\theta$ through $\pi$ results in a phase transition. This
behavior inspired the well-known proposal that the Hall plateau transition is
described by $N\to 0$ replica limit of the $U(2N)/U(N)\times U(N)$
sigma model with a $\theta$ term \cite{Pruisken}.

The theta angle need not be a tunable parameter.  If there is a
symmetry of the model (like time-reversal or $CP$ invariance) under
which the theta term is odd, then $\theta$ is fixed at zero or
$\pi$. Moreover, in certain sigma models such as $SU(N)/SO(N)$, the
winding number can only be $0$ or $1$, so $\theta$ can be only zero or
$\pi$. As we will discuss below, in such situations $\theta$ is
determined by the underlying disordered model. When $\theta=\pi$, the
$SU(N)/SO(N)$ model has a stable low-energy fixed points: it flows to
$SU(N)_1$ \cite{PF}. Thus the fixed value of $\theta$ determines
whether the model is gapless or not.

There is another way to modify some two-dimensional sigma models to
obtain a stable low-energy fixed point. This is to add a
Wess-Zumino-Witten (WZW) term to the action \cite{WZ}.  In
four-dimensional particle physics, the WZW term for example describes
the $\pi^0 \to 2\gamma$ decay in low-energy effective field theory
\cite{WZ}.  Such terms arise when the fields take values in a group.
Consider a theory with global symmetry $H\times H$, where $H$ is any
simple Lie group (e.g.\ $SU(N), SO(N)$, or $Sp(2N)$).  If the vacuum
manifold of the theory is chosen such that the modes in the diagonal
``vector'' subgroup $H_V$ become gapped, the low-energy fields, $h$,
take values in the remaining space, $H\times H/H_V$.  This space is
isomorphic to $H$ itself, and is called the axial part of the original
symmetry group.  For example, for $SU(N)$, $h$ would be an $N\times N$
unitary matrix with determinant one.

Without the WZW term, a sigma model where the fields take values in a
simple Lie group is called a principal chiral model. In terms of the
coupling, $g$ (the inverse stiffness), its action is
\begin{equation}
S_{PCM} = \frac{1}{g} \int dx\,dy\ 
\hbox{Tr}\left[\partial_\mu h\,  \partial^\mu h^{-1}
\right].
\label{pcm}
\end{equation}
We assume that the fields all fall to a constant at spatial infinity,
so that the two-dimensional spacetime can be treated as a sphere.  To
write the WZW term, one needs to extend the fields $h(x,y)$ on
the sphere to fields $h(x,y,z)$ on a ball which has the original
sphere as a boundary. The fields inside the ball are defined so that
$h$ at the center is the identity matrix, while $h$ on the boundary is the
original $h(x,y)$. It is possible to find a continuous deformation of
$h(x,y)$ to the identity because $\pi_2(H)=0$ for any simple Lie group.
Then the WZW term is $k\Gamma$, where
\begin{equation}
\Gamma= \frac{\epsilon_{ijk}}{24\pi}
\int dx dy dz\, \hbox{Tr}\left[(h^{-1}\partial_i h) 
(h^{-1}\partial_j h) (h^{-1}\partial_k h) \right].
\label{wzw}
\end{equation}
The coefficient $k$ is known as the level, and for compact groups must
be an integer because the different possible extensions of $h(x,y)$ to
the ball yield a possible ambiguity of $2\pi$ times an integer in
$\Gamma$.  Unlike the $\theta$ term, the WZW term does change the
equations of motion, but only by terms involving $h(x,y)$: the
variation of the integrand is a total derivative in $z$.

The sigma model with WZW term has a stable fixed point at $g=16\pi/k$
\cite{WZW}, so the model is critical and the
quasiparticles are gapless. 
 The corresponding con\-for\-mal field theory is
known as the $H_k$ WZW model \cite{KZ}.
The WZW term is invariant under discrete parity transformations (e.g.\
$x\to -x$, $y\to y$) provided $h$ is simultaneously transformed via
$h\to h^{-1}$.
In other words, the WZW term in a parity-invariant theory
involves pseudoscalars.

The WZW and theta terms are closely related.  For example, if one adds
a term $\lambda(\hbox{Tr}\, h)^2$ to the $SU(2)_k$ WZW action, at
$\lambda$ large the low-energy modes live on the sphere, i.e.\ the
$SU(2)/O(2)$ sigma model. One can check explicitly that the WZW term
then turns in to the theta term with $\theta=k\pi$
\cite{Haldane}. Moreover, in sigma models with only winding number $0$
or $1$, the topological term can be written in
the form of the WZW term (\ref{wzw})
\cite{WZW}.

Given the above considerations, it is important to understand how
a WZW term arises in
the replica formulation of two-dimensional models with disorder.  We
illustrate the appearance first by considering a simple theory where
such a term appears, related to models studied in
\cite{Gade,Senthil2}.  We start with a system of spin-polarized
fermions with a triplet p-wave type pairing:
\begin{equation}\label{eiii}
H_0 = \sum_k 
\epsilon_k c^\dagger_k c_k+(\Delta_k c^\dagger_k 
c^\dagger_{-k} + {\rm h.c.}),
\end{equation}
where $\Delta_k \sim {v_\Delta \over 2}k_x$.
We study disorder weak enough to maintain some notion of the Fermi surface.
The low energy excitations of the fermions are then found about two
nodes positioned on the $k_y$-axis at $K_{\pm} = (0,\pm K)$.  We
then linearize the theory about these
nodes via $\epsilon_{K_\pm + q} = q_y v_F$
and $c \sim c_1 \exp (iK_+ x) + c_2 \exp (iK_- x)$.  Introducing the
spinor $\psi^\dagger = (c^\dagger_{1},c_{2})$, we can write 
$H$ as
\begin{equation}\label{eiv}
H = \int d^2x~ \psi^\dagger (iv_F \tau_z\partial_y
+ i v_\Delta \tau_x \partial_x)\psi .
\end{equation}
The Pauli matrices $\tau_i$ act in the particle-hole space of the
spinors.  $H$ is precisely the Hamiltonian of a single Dirac fermion
in $2+1$ dimensions. To obtain the sigma model description of this
Hamiltonian, we fix the energy, $\omega$, at which we work, 
and describe the theory in terms of an
action of a two-dimensional Euclidean field theory:
\begin{equation}\label{ev}
S_0 = \int d^2x~ \psi^\dagger(iv_F\tau_z\partial_y
+ i v_\Delta \tau_x\partial_x - i\omega \tau_z) \psi.
\end{equation}
With this action we can compute correlators of the form 
$\langle 1/(H-\omega )\rangle$.

We include on-site disorder by adding a term,
\begin{equation}\label{evi}
H_{\rm disorder} = t (c^\dagger_1 c_1 + c^\dagger_2 c_2),
\end{equation}
to the Hamiltonian. Here $t$ is a random variable with variance $\langle
t(x) t(y)\rangle = (2u)^{-1} \delta (x-y)$.  To compute correlators with
disorder we replicate the theory, $\psi \rightarrow \psi_k$.
Disorder
is then easily averaged over leaving a set of quartic terms, $S_{\rm
disorder} = -{1\over u}
(\psi^\dagger_k\tau^z\psi_k)(\psi^\dagger_l\tau^z\psi_l)$. Reorganizing
the fields via $\tilde\psi^\dagger \equiv
(\psi^\dagger_R,\psi^\dagger_L) =
\psi^\dagger\tau^z\exp(i\pi\tau^x/4)$ and $\tilde\psi \equiv (\psi_R,
\psi_L) = \exp(-i\pi\tau^x/4)\psi$, gives the action (at $\omega = 0$) to be:
\begin{equation}
S = \int d^2 x i\tilde\psi^\dagger_k(v_F\partial_y \!-\!
v_\Delta\tau^z\partial_x)\tilde\psi_k
\!-\! {1\over u}(\tilde\psi^\dagger_k\tilde\psi_k)
(\tilde\psi^\dagger_l\tilde\psi_l).
\end{equation}
This theory is invariant under the group $U(N)_L\times
U(N)_R$ transforming $\tilde{\psi}\to \frac{1}{2}((1+\tau^z)U_L + (1-\tau^z)
U_R)\tilde{\psi}$.  Adding disorder to the Cooper-pairing term results in
another four-fermion term, but preserves this symmetry. As long as the
symmetry structure is unchanged, the low-energy sigma model should be
the same.

When $\omega=0$, this action is equivalent to a massless Dirac fermion
in a random magnetic field; the chiral symmetry prevents a mass term
from appearing.  This model can be solved by bosonization \cite{LFGS}.
Free fermions are equivalent to the WZW model $SU(N)_1\times U(1)$
\cite{WZW}.  The four-fermion term is then simply expressed as
$\int (J_L+ J_R)^2$, where $J_L$ and $J_R$ are the $U(1)$ currents,
$J_L=(\psi^{\dagger k}_L \psi_L^k )$ and $J_R=(\psi^{\dagger k}_R
\psi_R^k )$.  These $U(1)$ currents can be expressed in terms of a
free boson with $j_L=i\partial_L \phi$ and $j_R=-i\partial_R\phi$.
The four-fermion coupling then merely determines the boson
radius.  Therefore, the model reduces to pure
$SU(N)_1$ together with a decoupled free boson,
and hence is critical.

To see directly how the WZW term appears, it is useful to rederive
this result from the sigma model approach \cite{chetan}.  We assume
there is an energy scale where some fermion bilinear gets an
expectation value. Formally speaking, we introduce a Hubbard-Stratonovich
matrix field, $M_{kl}$, to factor the four-fermion term:
\begin{equation}\label{eviii}
S = S_0 - \int d^2x \big( {u \over 4}\hbox{Tr}(M^2)
+i (\tilde\psi^\dagger_k M_{kl}\tilde\psi_l)\big),
\end{equation}
where $M$ is hermitian. Under $U(N)_L\times U(N)_R$, $M \to UMU^\dagger$. 
The WZW term appears in low-energy effective theories describing
saddle points where $M$ is off-diagonal, e.g. \ $M_{LL} = M_{RR} = 0$,
but $M_{LR} = M^\dagger_{RL}\propto I$, the identity.  The
diagonal subgroup $U(N)_V$ leaves this saddle point invariant, and so the
low-energy modes $T=M_{RL}$ take values in $U(N)_L\times U(N)_R/U(N)_V
\approx U(N)$. Note that under parity, $T^\dagger\leftrightarrow
T$.
Focusing solely upon these modes allows us to write
\begin{equation}\label{eix}
S = S_0 - \int d^2x \big({u \over 2}\hbox{Tr}(TT^\dagger)
+i (\tilde\psi^\dagger_R T\tilde\psi_L
+ \tilde\psi^\dagger_L T^\dagger\tilde\psi_R)\big).
\end{equation}
Integrating out the fermions leaves an effective action for the
bosonic field $T$, which can be expanded in powers of the momentum
over the expectation value of $T$.  At low energy, one can safely
neglect four-derivative terms and higher, and one easily obtains the
kinetic term (\ref{pcm}) for $T$.  However, a less-obvious
term also appears.  The case of interest was treated in \cite{DF}.  To perform a consistent low-energy
expansion, one must change field variables. This results in a Jacobian
in the path integral \cite{Fuj}, which is precisely the WZW term.
Writing $T=he^{i\phi}$, where det($h)=1$, we obtain here the WZW term
(\ref{wzw}) of the $SU(N)$ field $h$, with level $k=1$. The extra
$U(1)$ field $\phi$ in the sigma model is a theory of a
decoupled boson.

Thus in the sigma model approach we recover the critical line
described by $SU(N)_1 \times U(1)$ conformal field theory in the limit
$N\to 0$. This critical line is a part of the critical space of
the Gade hopping model; the full space is
larger because the $SU(N)$ coupling does not flow when $N=0$
\cite{Gade,andreas}.
Moreover, this same
CFT also applies to a time-reversal symmetric
version of the hopping model, which is described by a $U(N)/O(N)$
sigma model. When $\theta=\pi$, this sigma model flows to the same
$SU(N)_1\times U(1)$ conformal field theory \cite{PF}.  We can see how
the two are related by perturbing the $U(N)$ WZW model action by a
term which gives a mass to the modes outside a $U(N)/O(N)$ subspace
(the subspace corresponds to symmetric $U(N)$ matrices).  As
with the $N=2$ case mentioned above, the low energy limit of the
perturbed $U(N)$ WZW model reduces to the $U(N)/O(N)$ sigma model with
$\theta=k\pi$. Thus the Gade model with and without time reversal
invariance provides a concrete realization of the equivalence of theta
and WZW terms in disordered systems.

In no sense does the WZW term arise in the above models as a result of
fine tuning: it must appear when there is a chiral anomaly. In
the model treated above, the fermions have a $U(N)_L \times U(N)_R$
symmetry.  As is well known, chiral symmetries involving fermions are
frequently anomalous, so that the Noether currents do not all remain
conserved in the quantum theory. For
massless fermions in $1+1$ dimensions, this was shown in detail in
\cite{CGJ,WZW}.  The WZW term is the effect of the anomaly on the
low-energy theory. Even though the fermions effectively become massive
when $T$ gets an expectation value, their presence still has an effect
on the low-energy theory, even if this mass is arbitrarily large.
This violation of decoupling happens because the chiral anomaly must
be present in the low-energy theory. In other words, the anomaly
coefficient does not renormalize. This follows from an argument known
as 't Hooft anomaly matching \cite{Hooft}. One imagines weakly gauging
the anomalous symmetry (in our case the axial $SU(N)$). It is not
possible to gauge an anomalous symmetry in a renormalizable theory,
but one can add otherwise non-interacting massless chiral fermions to
cancel the anomaly. Adding these spectator fermions ensures that the
appropriate Ward identities are obeyed and the symmetry can be
gauged. In the low-energy effective theory, the Ward identities must
still be obeyed and the theory must remain anomaly-free. Because the
massless spectator fermions are still present in the low-energy
theory, there must be a term in the low-energy action which cancels
the anomaly from the spectators. This is the WZW term.

We now turn to a case of considerable interest, that of $d_{x^2-y^2}$
superconductors in the presence of disorder \cite{NTW,Senthil}.
We now study fermions with spin symmetry, so that 
$c,c^\dagger$ carry a node index, a spin
index, and a replica index.  We label the nodes $(1,\pm)$ for the pair
near the wavevectors $k=\pm(\pi/2,\pi/2)$ and $(2,\pm)$ for the pair
near $k=\pm(-\pi/2,\pi/2)$.  It is convenient to group particles and
holes together.  Thus define four doublets,
$\psi^{j}_{a\alpha}=(c^j_{a\alpha},i(\sigma^y
s^x c^{j\dagger})_{a\alpha})$, $a=\pm, \alpha = \uparrow/\downarrow$,
for each pair of nodes $j=1,2$.  The Pauli matrices
$\sigma^a$ and $s^a$ act
upon the spin/$\pm$ node indices respectively.  If we rotate the 
system in the $ab$ plane by $45^o$, the
linearized Hamiltonian for a superconductor with time-reversal
and spin symmetry is 
\begin{equation}
H=i\overline{\psi}^1(v_F s^z\tau^z\partial_x +
v_\Delta s^z\tau^x\partial_y)\psi^1 + 
(x,1\leftrightarrow y,2),
\end{equation}
where the Pauli matrix $\tau$ acts in particle/hole space and
$\overline{\psi}^j \equiv (s^x\sigma^y\tau^y
\psi^j)^T$.  After adding disorder, the replicated Hamiltonian
is invariant under the group $Sp(2N)_L\times Sp(2N)_R$
\cite{Senthil}. It sends $\psi^i \to U\psi^i$, where
$U=\frac{1}{2}\left[U_L(1+\tau^y) + U_R(1-\tau^y)\right],$ with $U_L$
and $U_R$ each elements of $Sp(2N)$ acting only upon spin and replica
indices.  An element $Q$ of $Sp(2N)$ is an invertible $2N\times 2N$
real matrix obeying $Q^T\sigma^y Q = \sigma^y$.  With Gaussian on-site
disorder and $\omega=0$, the low-energy modes of the corresponding
action take values in $Sp(2N)_L\times Sp(2N)_R/Sp(2N)_V$.  The
low-energy action thus includes the term (\ref{pcm}) with $h$ in
$Sp(2N)$ \cite{Senthil}.

This model also has a WZW term.  We first consider the case where
disorder does not couple the pairs of nodes. Once disorder has been
averaged over and the quartic terms have been factorized, the fermions
interact with hermitian matrix fields, $M^{(1)}$ and $M^{(2)}$, via terms
of the form
\begin{equation}
\bar\psi^1_{\alpha} M^{(1)}_{\alpha\beta}\psi^1_{\beta} + \bar\psi^2_{\alpha}
M^{(2)}_{\alpha\beta}\psi^2_{\beta},
\label{super}
\end{equation}
as before.  At a saddle point with non-zero values of the off-diagonal
elements, the argument of \cite{DF} shows that we will again obtain
level 1 WZW theories for $M^{(1)}$ and for $M^{(2)}$, in this case
$Sp(2N)_1$.

We make this explicit by showing how the spin
degrees of freedom expand the previous $U(N)$ symmetry to that of
$Sp(2N)$. $Sp(2N)$ has a $U(N)$ subgroup consisting of orthogonal
real matrices of the form
$$\pmatrix{A&B\cr -B &A\cr},$$ where we put the up spins in the first
$N$ components of $\psi$, and the down spins in the second half.  Thus
the $U(N)$ subgroup acts on the combinations, $\psi_{\uparrow} \pm
i\psi_{\downarrow}$, with the matrices $A\mp i B$. In group-theoretic
language, the $2N$ dimensional representation of $Sp(2N)$ decomposes
into the $N+\overline{N}$ representation of $U(N)$.  We can examine
the WZW model describing the low-energy excitations which lie in the
$U(N)$ subspace. It must be exactly the same as that treated above,
with the exception that there are now two species of ``spinless''
fermions $\psi_{\uparrow}\pm i\psi_{\downarrow}$. This results in two
WZW terms, one with $h$ and one with $h^*$.  As the WZW term is
invariant under complex conjugation, the two contributions are
identical, and so add. This theory is a $U(N)_2$ WZW theory (as was
also derived using bosonization \cite{NTW}; the corresponding result
in the supersymmetric formulation was derived in \cite{zirn}).  The
level of the WZW term in the full $Sp(2N)$ theory is determined by the
embedding: when an $H_l$ theory is embedded in $G_k$, the levels obey
$kr=l$, where $r$ is a group-theory factor called the index of the
embedding of $H$ into $G$. For the embedding of $U(N)$ into $Sp(2N)$,
$r=2$. Thus we have shown that the low-energy theory for (\ref{super})
is given by an $Sp(2N)_1$ WZW model for each pair of nodes.

When a WZW term is present, the quasi-particles of the
spin/time-reversal invariant superconductor need not be localized.
Specifically, the coupling constant of the $Sp(2N)_k$ theory,
inversely related to the spin conductance, now flows to a fixed value
(provided its bare value is sufficiently small), as opposed to
becoming arbitrarily large.  Thus with a WZW term such models are 
no longer spin insulators.
The $Sp(2N)_k$ WZW models seem to be well
behaved as $N\to 0$. The dimensions of the fundamental operators are
given by $m(2-m)/(2k+2)$, where $m$ is a positive integer. To compute
the density of states at a finite energy, $\omega$, we add a term of
the form $\omega {\rm Tr}(h+h^\dagger)$ to the action. The density of
the states is then given by $\rho (\omega ) \sim \langle {\rm Tr}\,
(h+h^\dagger)\rangle$.  This field $\hbox{tr }h$ has dimension $\Delta
= 1/(2k+2)$, and a simple scaling argument yields $\rho (\omega ) \sim
\omega^{\Delta/(2-\Delta)} = \omega^{1/(4k+3)}$. This result agrees
with a previous computation\cite{NTW} when $k=1$.

When the two pairs of nodes are not coupled by the disorder, the
low-energy theory is therefore described by {\it two} $Sp(2N)_1$ WZW
models, which have symmetry currents, $J_{L1,2},J_{R1,2}$.
Once the pairs of nodes are coupled, there are three
possibilities. The first
is that the two $Sp(2N)_L\times Sp(2N)_R$ symmetries are preserved, so
the low-energy theory remains two $Sp(2N)_1$ WZW models.  The second
is that only simultaneous chiral transformations on the two pairs
remain a symmetry. In the decoupled theory, this symmetry is generated
by $J_{L1} + J_{L2}$ and $J_{R1} + J_{R2}$.  It has level $2$ (in the
language used above, the index of the embedding is $2$).  Once the
nodes are coupled, the generators may be deformed away from $J_{L1} +
J_{L2}$ and $J_{R1} + J_{R2}$, but the level cannot change.  This
occurs for the same reasons that anomalies remain in the low-energy
theory -- the anomaly coefficient cannot renormalize.  Thus
the low-energy behavior of this model
is described by an $Sp(2N)_2$ WZW model. If one
modifies the Hamiltonian so that the two pairs of nodes are no longer
identical, then the result is the same as long as the chiral symmetry
is not explicitly broken: the chiral anomaly does not change under
perturbation.

The third possibility is that the anomalies cancel between the two
theories. This is in fact what happens in $d_{x^2-y^2}$
superconductors \cite{ASZ}. 
Consider two coupled WZW
models with fields $h_1$ and $h_2$ and action
$$S_{PCM}(h_1) + \Gamma(h_1) + S_{PCM}(h_2) + \Gamma(h_2) +
\lambda\hbox{Tr}(h_1 h_2 + h.c.).$$ The global symmetries of the
decoupled model are $h_{i} \rightarrow U_{Li}h_{i}U_{Ri}$.  In the
coupled model this global symmetry is broken to $h_1 \rightarrow
U_{L1}h_1U^\dagger_{L2}$, $h_2 \rightarrow U_{L2}h_2U^\dagger_{L1}$,
appropriate to the underlying lattice symmetry of the d-wave
superconductor.  These symmetries are anomaly free; the axial
anomalies of the two nodes cancel upon coupling.  This can be seen
roughly in that the coupling $\lambda$ is relevant; in the strong
coupling limit $h_1 = h_2^\dagger$ (i.e.  $\Gamma (h_1) = -\Gamma
(h_2) $) and so the WZW terms cancel.  This omits terms that might
arise in integrating out massive degrees of freedom. This
possibility cannot be ignored, because we know this is precisely how a WZW
term arises in \cite{DF}. However, this argument can be made more
precise in the case when $h_1$ and $h_2$ are elements of $O(N)$
instead of $Sp(2N)$.  Indeed, then one can fermionize one of the two
WZW models, writing $(h^2)_{ab} =\bar\psi_b\psi_a$, with the resulting
action
$$S_{PCM}(h_1) + \Gamma(h_1) + \bar\psi\gamma^\mu\partial_\mu\psi +
\lambda(\bar\psi h_1 \psi + h.c.).$$ 
For $\lambda$ large, we can integrate out these fermions 
according to \cite{DF}
as before.  In doing so, a WZW term for $h_1$ is induced
which precisely cancels the original WZW term for
$h_1$.  If we had instead coupled the theories via a term $\hbox{Tr}(h_1
h_2^\dagger + h.c.)$, the two terms would have added, and we would
have recovered the level two theory discussed above.  
This cancellation occurs
even if the two theories have differing Fermi velocities.  Including
Fermi velocities directly in the action (easily done
through dimensional analysis)
shows that the terms $S_{PCM}$ are affected while the WZW 
terms $\Gamma$ are not.
Thus the cancellation between the two WZW terms can proceed as above.

A d-wave superconductor with spin rotational but broken time reversal
invariance is realized when the gap wave function is $\Delta \sim
d_{x^2-y^2}+id_{xy}$.  This breaks the $Sp(2N)\times Sp(2N)$ symmetry
explicitly, and the appropriate sigma model in the replica formulation
for this theory is $Sp(2N)/U(N)$ with a theta term \cite{Senthil,brad}.
Because the time-reversal invariance is broken, the coefficient of
$\theta$ is not constrained by symmetry, but at $\theta=\pi$ one
expects a critical point in the same fashion as \cite{Pruisken}. It is
conjectured in \cite{PF} that at $\theta=\pi$, this model flows to the
$Sp(2N)_1$ conformal field theory. By utilizing the supergroup
formulation of the corresponding network model (spin
quantum Hall effect), it was shown that the critical point point of
the disordered model is equivalent to classical percolation
\cite{ilya}. Thus the replica limit $N\to 0$ of $Sp(2N)_1$ should be
equivalent to the critical point in the spin quantum Hall effect. The
exponents computed above do indeed agree with those computed in
\cite{ilya}, for example $\rho(E) \propto E^{1/7}$.  We again see a
feature that appeared in the $U(N)$ class of problems: related models
seemingly in two different universality classes are in the same one,
once the WZW terms and $\theta$ terms are specified.

We have shown how stable critical points can arise in a number of
replica sigma models.  This shows that a number of disordered symmetry
classes in two dimensions \cite{zirn} allow further refinement.  In
particular, in models with a chiral symmetry, one must compute the
anomaly in the original underlying short-distance replica (or
supergroup) model to determine the WZW term in the low-energy
model.  Likewise, in sigma models with ${\bf Z}_2$ winding numbers,
one must determine whether $\theta=0$ or $\pi$.

\bigskip
We would like to thank K.~Intriligator, A.~Ludwig, C.~Nayak, T.~Senthil,
B.~Simons and M.~Zirnbauer for
extremely helpful conversations.
This work was supported by a DOE OJI
Award, a Sloan Foundation Fellowship, and by NSF grant DMR-9802813.


\begin{references}
\bibitem{SW} L. Sch\"afer and F. Wegner, Z. Phys. B38 (1980) 113
\bibitem{Gade} R. Gade, Nucl. Phys. B398 (1993) 499; R. Gade and F. Wegner,
Nucl. Phys. B360 (1991) 213
\bibitem{ZZ} A.B. Zamolodchikov and Al.B. Zamolodchikov, 
Nucl. Phys. B 379 (1992) 602
\bibitem{Haldane} For a review see I. Affleck in {\it Fields, Strings
and Critical Phenomena} (North-Holland 1988).
\bibitem{Pruisken} A. Pruisken, Nucl. Phys. B235 (1984) 277 
\bibitem{PF} P. Fendley, ``Integrable sigma models with $\theta=\pi$'',
in preparation.
\bibitem{WZ} J. Wess and B. Zumino Phys. Lett. 37B (1971) 95;
E. Witten, Nucl. Phys. B223 (1983) 422
\bibitem{WZW} E.~Witten, Comm.~Math.~Phys.~92 (1994) 455
\bibitem{KZ} V.~Knizhnik, A.~Zamolodchikov, Nucl.~Phys. B247 (1984) 83
\bibitem{Senthil2} T. Senthil and M.P.A. Fisher, Phys. Rev. B 61 (2000) 9690, cond-mat/9906290.
\bibitem{chetan} We thank Chetan Nayak for sharing his notes on 
fermionic replicas for \cite{Gade} with us.
\bibitem{LFGS} A. Ludwig, M.P.A. Fisher, R. Shankar and G. Grinstein,
Phys. Rev. B50 (1994) 7526. We thank Andreas Ludwig for pointing this
out to us.
\bibitem{andreas} S.~Guruswamy, A.~LeClair and A.~Ludwig, cond-mat/9909143
\bibitem{DF} E. D'Hoker and E. Farhi, Nucl. Phys. B248 (1984)
59; E. D'Hoker and D. Gagn\'e, Nucl. Phys. B467 (1996) 272,
hep-th/9508131
\bibitem{Fuj} K. Fujikawa, Phys. Rev. D29 (1984) 285
\bibitem{Hooft} G. 't Hooft, in {\it Recent Developments in Gauge
Theory} (Plenum 1980).
\bibitem{CGJ} S. Coleman, D. Gross and R. Jackiw, Phys. Rev. 180 (1969) 1359
\bibitem{NTW} A.A. Nersesyan, A.M. Tsvelik, F. Wenger,
Nucl. Phys. B438 (1995) 561; cond-mat/9401026
\bibitem{Senthil} T. Senthil, L. Balents, C. Nayak and M.P.A. Fisher,
Phys. Rev. Lett. 81 (1998) 4704, cond-mat/9808001; T. Senthil and
M.P.A. Fisher, Phys. Rev. B60 (1999) 6893, cond-mat/9810238
\bibitem{zirn} M. Bocquet, D. Serban and M.R. Zirnbauer, cond-mat/9910480
\bibitem{ASZ} A. Altland, B. Simons and M.R. Zirnbauer, to appear;
A. Ludwig and T. Senthil, unpublished; T. Fukui, cond-mat/0002002
\bibitem{brad} V. Kagalovsky, B. Horovitz, Y.Avishai, J. T. Chalker,
Phys. Rev. Lett. 82 (1999) 3516, 
cond-mat/9812155; T. Senthil, J. B. Marston,
M.P.A. Fisher, Phys. Rev. B 60 (1999) 4245, cond-mat/9902062
\bibitem{ilya} 
I.A. Gruzberg, A. Ludwig, N. Read, Phys. Rev. Lett. 82 (1999) 4524,
cond-mat/9902063;  D. Bernard and A. LeClair, cond-mat/0003075;
J. Cardy and J. Chalker, to appear
\bibitem{zirn} A. Altland and M. Zirnbauer, Phys. Rev. B55 (1998) 1142,
cond-mat/9602137; M. Zirnbauer, J. Math. Phys. 37 (1996) 4986, math-ph/9808012

\end{references}
\end{document}